# Toward all-optical unsupervised Hebbian learning in deep photonic neuromorphic networks

**Authors:** Xi Li[1], Disha Biswas[2], Peng Zhou[2], Wesley H. Brigner[2], Anna Capuano[3], Joseph S. Friedman[2*], Qing Gu[1,4,*]

**Affiliations:**

[1]Department of Electrical and Computer Engineering, North Carolina State University, Raleigh, NC 27695, USA

[2]Department of Electrical and Computer Engineering, The University of Texas at Dallas, Richardson, TX 75080, USA

[3]Department of Materials Science and Engineering, North Carolina State University, Raleigh, NC 27695, USA

[4]Department of Physics, North Carolina State University, Raleigh, NC 27695, USA

*Corresponding authors. Email: qgu3@ncsu.edu (Q.G), joseph.friedman@utdallas.edu (J.S.F)

**Abstract:** We propose a deep photonic neuromorphic network (PNN) architecture based on phase-change material (PCM) synapses and local optical feedback for online, unsupervised Hebbian learning. The proposed architecture combines optical vector-matrix multiplication, non-volatile PCM synaptic weighting, and local coincidence-driven synaptic adaptation within a multilayer photonic crossbar framework compatible with photonic integrated circuits. Unlike conventional PNNs that rely on externally computed gradients, repeated optical-electrical-optical conversions, or global backpropagation, the proposed framework employs local Hebbian learning governed directly by correlated pre- and post-synaptic optical activity. To investigate the feasibility of the proposed learning mechanism, we implemented the PNN design using fiber-optic components, programmable variable optical attenuators, and real-time software control that incorporates PCM thermal dynamics. Supervised and unsupervised learning behaviors were experimentally evaluated under both offline and online learning conditions using representative image-recognition tasks. The experimental results demonstrate adaptive synaptic evolution, successful optical inference, and autonomous pattern encoding through local Hebbian learning under realistic fiber-optic hardware conditions. These results establish a pathway toward future integrated photonic neuromorphic systems capable of scalable and energy-efficient online Hebbian learning.

## 1. Introduction

In the era of big data and artificial intelligence (AI), demand for high-performance, energy-efficient computing hardware has grown rapidly. Neuromorphic computing, inspired by the architecture and functionality of biological neural systems, offers a promising paradigm for overcoming the data-transfer bottlenecks and energy limitations of conventional von Neumann architectures. By utilizing hardware that replicates the functionality of synapses and neurons and co-locates memory and computation, neuromorphic systems enable massively parallel, energy-efficient, and event-driven information processing with reduced latency and improved scalability [1–4].

While electronic neuromorphic processors have made significant progress [5–11], photonic neuromorphic networks (PNNs) have emerged as an attractive alternative, because optical signals inherently provide high bandwidth, low latency, and parallel information processing [2,12]. Recent advances in integrated photonics have enabled a wide range of PNN implementations, including photonic tensor cores [13], a coherent Mach-Zehnder interferometer-mesh PNN [14], optoelectronic PNN [15], and diffractive optical neural networks [16,17]. Despite these advances, most integrated solutions still rely heavily on optical-electrical-optical (O-E-O) conversions, externally computed gradients, offline training, or electronically assisted weight updates, which limit the achievable speed and energy efficiency of photonic learning systems [18].

In parallel, several free-space optical neural network platforms have demonstrated fully optical forward and backward propagation for training neural networks [19]. These approaches leverage the intrinsic parallelism of free-space optics to implement optical backpropagation and optical gradient computation without intermediate electronic processing. However, free-space optical systems generally face challenges in compact scalability, environmental stability, and dense on-chip integration. In contrast, integrated PNNs based on photonic integrated circuits (PICs) provide a promising pathway toward scalable, robust, and energy-efficient neuromorphic hardware suitable for practical deployment. Realizing online learning directly within integrated photonic hardware, therefore, remains an important open challenge.

Another major challenge lies in implementing online learning directly within photonic hardware. Most demonstrated PNNs perform inference optically while training is executed offline using digital computers, followed by weight transfer to the photonic hardware [20,21]. Alternative approaches based on direct feedback alignment and physical neural network training have demonstrated in-situ or hardware-aware optimization of photonic systems [21–24]. In particular, Momeni et. al. demonstrated backpropagation-free training of deep physical neural networks using experimentally measured physical forward propagation combined with digitally computed updates [25]. However, these approaches still rely on global optimization procedures, externally computed learning signals, or electronic control loops rather than local synaptic plasticity mechanisms inspired by biological neural systems.

In contrast, unsupervised Hebbian learning relies on local correlations between pre- and post-synaptic activity, and therefore, provides a promising route toward scalable and hardware-efficient photonic learning systems [26–28]. Previous work has explored spike-timing-dependent plasticity and self-learning photonic synapses using phase-change materials (PCMs) [29–31], but these demonstrations were limited to device-level and shallow PNN demonstrations, and did not address deep multi-layer photonic architectures with online unsupervised learning. Furthermore, many existing photonic synapse implementations, including Mach-Zehnder interferometers, micro-ring resonators, and homodyne-based architectures, require continuous electrical biasing or external supervision, limiting scalability and energy efficiency [14,31–33].

To this end, PCMs offer a promising solution for realizing non-volatile photonic synapses. PCMs exhibit large and reversible optical-property contrast between amorphous and crystalline states, while maintaining long-term state retention without static power consumption [34–38]. In addition, PCM devices can support gradual and reversible optical weight modulation, making them attractive candidates for neuromorphic photonic hardware [39–41]. When combined with local optical feedback, PCM-based synapses offer a potential pathway toward fully integrated photonic learning systems that avoid repeated O-E-O conversions during both inference and learning. Feldmann et al. previously demonstrated an integrated PCM-neuron and PCM-synapse shallow spiking PNN with supervised pattern recognition and local unsupervised synaptic plasticity [30].

Our work builds on this direction from a different architectural perspective by proposing a multilayer crossbar-style PNN with local feedback and state-dependent PCM Hebbian updates, aimed at online unsupervised learning in a deep PNN rather than single-layer spiking-network operation. The proposed architecture employs optically controlled PCM synapses and PCM microring neurons to enable local synaptic updates via coincident pre- and post-synaptic optical signals. Unlike backpropagation-based photonic training approaches that require global error propagation and externally computed gradients, the proposed framework employs a local Hebbian learning rule that depends only on the presynaptic input, postsynaptic feedback, and current synaptic state. We further experimentally validate the proposed learning framework using a fiber-optic PNN that implements online unsupervised learning and real-time adaptive weight updates. Using a non-trivial letter-recognition task, we demonstrate both supervised and unsupervised online learning in a multilayer PNN platform. These results establish a pathway toward future integrated PNNs capable of scalable and energy-efficient online learning.

**Table 1** Summary of state-of-the-art PNNs. Here, "global optimization" refers to methods such as supervised backpropagation, feedback alignment, and physical neural network training; "local synaptic update" refers to Hebbian learning.

| Platform | Learning method | Scale | Local synaptic update | Non-volatile weights | Online adaptation | PIC-compatible architecture |
|---|---|---|---|---|---|---|
| Integrated nanophotonic circuit [14] | Offline backpropagation | Deep | | | | ✓ |
| Free-space diffractive optics [16] | Offline deep learning | Shallow | | | | |
| Integrated PCM photonic synapses [30] | Self-learning/ photonic plasticity | Shallow | ✓ | ✓ | ✓ | ✓ |
| Physical neural networks [21] | In-situ backpropagation | Deep | | | ✓ | Partially |
| Physical neural networks [25] | Backpropagation-free global optimization | Deep | | | ✓ | Partially |
| Integrated photonic memory core [31] | Electrically programmed in-memory computing | Shallow | | ✓ | Limited | ✓ |
| Free-space optical neural network [19] | End-to-end optical backpropagation | Deep | | | ✓ | |
| Unsupervised PNN (this work) | Local Hebbian learning with optical feedback | Deep | ✓ | Simulated PCM synapses | ✓ | Proposed |

## 2. Proposed deep PNN architecture

We propose a deep PNN architecture based on non-volatile PCM synapses and local optical feedback for online unsupervised Hebbian learning. The proposed architecture is designed for future implementation on an integrated photonic platform, combining optical vector-matrix multiplication, nonlinear optical neuron activation, and local synaptic weight adaptation within a multilayer photonic crossbar network. Fig. 1 illustrates the conceptual architecture and its key functional components. The experimental fiber-optic setup used to validate the proposed learning framework is introduced later in Section 4.

As illustrated in Fig. 1a, the proposed deep PNN comprises cascaded photonic crossbar layers with programmable PCM synapses and neurons. Within each row of the crossbar, input optical signals are evenly distributed to the PCM optical synapses via directional couplers, whose coupling lengths are designed to ensure even power distribution among all synapses. During inference, the optical input vector is multiplied by the synaptic transmission matrix through passive optical propagation and attenuation within the PCM synapses. The weighted optical signals are subsequently integrated along each column and injected into optical neurons that generate nonlinear activation responses. A fraction of the neuron output is then routed to the subsequent network layer, while the rest is directed into a local optical feedback pathway that enables synaptic weight adaptation during learning. The color gradients and arrows in Fig. 1a indicate the distribution and accumulation of optical power through the network.

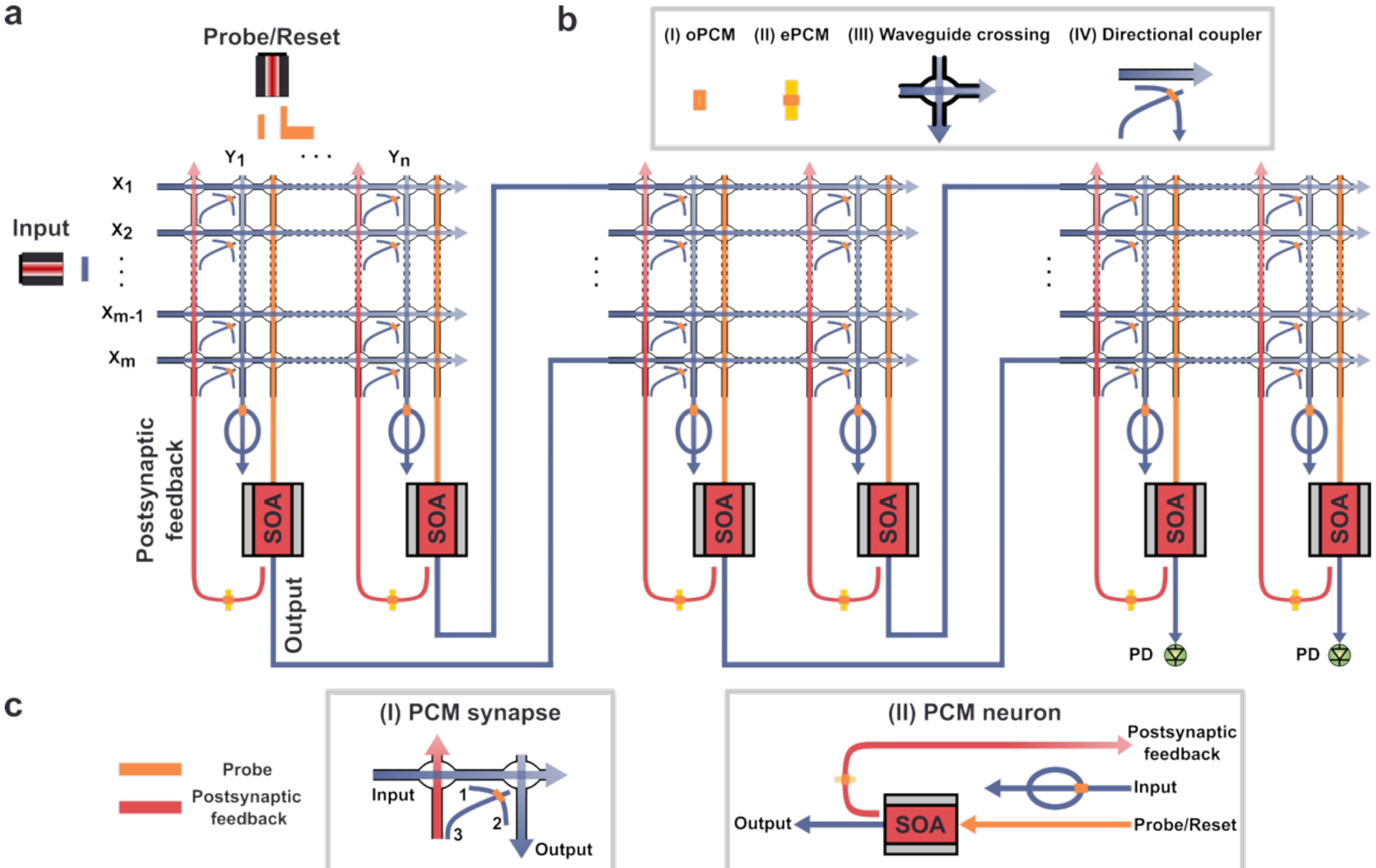

**Fig. 1** Proposed PCM-based deep PNN architecture. **(a)** Conceptual illustration of the proposed multilayer photonic crossbar architecture composed of non-volatile PCM synapses and optical neurons. Optical input vectors propagate horizontally through programmable PCM synapses and are summed vertically along postsynaptic waveguides. Arrows indicate the propagation of the optical signal and the weighted accumulation of optical signals within the network. A local optical feedback pathway routes a fraction of the postsynaptic optical signal back to synapses within the same column, enabling local Hebbian learning. The figure illustrates a proposed integrated photonic implementation; the experimental setup used for validation is introduced later in Section 4. **(b)** Key functional components of the proposed architecture. (I) Optically controlled PCM (oPCM) cells integrated at waveguide crossings implement programmable non-volatile synapses and PCM-based optical neurons. (II) Electrically controlled PCM (ePCM) switches activate or deactivate local feedback pathways during learning and inference. (III) Low-loss waveguide crossings enable compact row-column interconnections while minimizing optical crosstalk and insertion loss. (IV) Directional couplers with engineered coupling ratios distribute optical input power approximately uniformly across downstream synapses while maintaining optical intensities below PCM switching thresholds during inference. **(c)** Detailed operation of major functional blocks. (I) During learning, presynaptic input signals (blue) and postsynaptic feedback signals (red) overlap temporally and spatially within the PCM synapse located at the waveguide crossing. The resulting local optical energy deposition induces PCM phase transitions that modify synaptic transmission in accordance with a Hebbian learning rule. Although the feedback signal is distributed across the column, synaptic updates remain locally determined through coincidence detection between the presynaptic and postsynaptic optical signals within each PCM cell. (II) In the proposed PCM microring neuron, the integrated postsynaptic optical signal drives the phase transition of the PCM integrated atop the resonator, while a probe signal (orange) monitors the resonator transmission state. The resulting resonance shift and coupling modulation produce a nonlinear optical activation response that resembles a rectified linear unit (ReLU) function, thereby generating an output signal (blue) from the bus waveguide. A semiconductor optical amplifier (SOA) amplifies the neuron output before redistribution to subsequent network layers and local feedback pathways.

Fig. 1b(I-IV) details the optical synapses. They are implemented using optically controlled PCM (oPCM) cells integrated at waveguide crossings, as shown in Fig. 1b(I) and Fig. 1c(I). The PCM cells are designed to operate as non-volatile optical attenuators, with their transmission states encoding synaptic weights. PCMs exhibit large and reversible optical-property contrast between amorphous and crystalline states, including changes in both refractive index and absorption coefficient [34–36,42]. Consequently, the transmission of each waveguide crossing can be continuously tuned through partial phase transitions within the PCM cell, enabling analog optical weight modulation. Because the PCM states are non-volatile, the programmed synaptic weights can be retained without static power consumption during inference.

Within each row of the crossbar, the optical input signal is distributed to multiple synapses using directional couplers with engineered coupling ratios and coupling lengths, as illustrated in Fig. 1b(IV). The coupling coefficients are selected to ensure approximately equal optical power distribution among downstream synapses while maintaining optical intensities below the PCM phase-change threshold during inference. Waveguide crossings, illustrated in Fig. 1b(III), enable compact row-column interconnections while minimizing insertion loss and optical crosstalk. Inverse-designed crossings and low-loss passive photonic routing structures can be employed to improve scalability and network density.

The weighted optical signals from all rows are subsequently combined along each column and injected into optical neurons, illustrated in Fig. 1c(II). Here, the neuron consists of a PCM-integrated microring resonator operating near critical coupling. Initially, the PCM cell atop the microring is in its crystalline state, and the microring is designed such that the internal round-trip loss matches the coupling strength between the microring and the bus waveguide. Under this critical coupling, the probe signal is strongly attenuated because optical energy is trapped and dissipated within the microring neuron, corresponding to a low-transmission neuron output state.

When the integrated postsynaptic optical signal exceeds a threshold, optical absorption within the PCM cell of the microring neuron induces a phase transition from the crystalline to the amorphous state. This transition simultaneously modifies the resonator's effective refractive index and its optical absorption. As a result, the resonance wavelength shifts, and the coupling condition changes from critical coupling to the over-coupled regime. Consequently, the probe signal becomes increasingly transmitted through the bus waveguide rather than trapped within the microring resonator. The microring neuron’s output, therefore, increases monotonically with the degree of neuron PCM amorphization, which produces a nonlinear optical activation response resembling a rectified linear unit (ReLU) activation function [30]. After neuron activation, a semiconductor optical amplifier (SOA) amplifies the optical output signal before redistribution to subsequent network layers and local feedback pathways. To prepare the neuron for subsequent inference operations, a staircase-shaped optical reset pulse can be applied via an adjacent waveguide to gradually recrystallize the neuron PCM cell atop the microring and restore the microring neuron to its initial state.

During learning, the local optical feedback pathway illustrated in Fig. 1a and Fig. 1c(I) enables synaptic weight adaptation according to the Hebbian learning rule [28]. In inference mode, the feedback path is disabled using electrically controlled PCM switches (ePCM), illustrated in Fig. 1b(II). During learning, the feedback pathway is activated, and a fraction of the postsynaptic optical signal is distributed to all synapses within the same column. Importantly, although the postsynaptic feedback signal is shared across a column, synaptic weight updates remain locally determined because the phase transition within each PCM synapse depends jointly on the presynaptic input signal, the postsynaptic feedback signal, and the current synaptic state. Consequently, only synapses that experience sufficiently strong coincident pre- and post-synaptic optical activity undergo significant phase transitions and weight modifications. The shared-column feedback, therefore, functions analogously to a postsynaptic broadcast signal in neurobiology, while synaptic selectivity arises from local coincidence detection within each PCM cell, consistent with Hebbian learning principles.

Unlike conventional backpropagation-based photonic training approaches that rely on globally computed gradients, external optimization hardware, or repeated O-E-O conversions, the proposed framework implements local synaptic adaptation directly through optical signal interactions within the photonic crossbar. The detailed PCM phase-transition dynamics and synaptic weight evolution underlying this local Hebbian learning process are discussed in Section 3.

## 3. PCM synapse Hebbian learning model

The proposed deep PNN architecture leverages the non-volatile and reconfigurable optical properties of PCMs to implement local Hebbian synaptic plasticity directly in the photonic domain. In this work, germanium-antimony-tellurium (GST) is considered as the representative PCM because it exhibits large and reversible optical property contrast between its amorphous and crystalline phases while maintaining long-term non-volatile state retention [34–36,42]. In addition, GST supports gradual, reversible optical transmission modulation through partial phase transitions, enabling analog synaptic weight storage and update operations suitable for PNNs.

The synaptic weight evolution is governed by thermally induced phase transitions within the GST cell. GST can exhibit two stable phases with distinct optical properties: a crystalline phase with higher absorption and an amorphous phase with lower absorption. Local optical excitation modifies the relative fractions of these phases via temperature-dependent amorphization and crystallization, thereby altering the PCM synapse's optical transmission. Partial phase transitions allow intermediate crystalline-amorphous configurations to represent analog synaptic states, while the non-volatile nature of GST enables these states to be retained without continuous power consumption. Therefore, GST provides a reversible, analog, and non-volatile optical storage mechanism suitable for photonic neuromorphic systems.

Fig. 2a illustrates the proposed Hebbian learning mechanism within an individual PCM synapse. During learning, presynaptic input signals and postsynaptic feedback signals overlap temporally and spatially within the PCM-integrated waveguide crossing synapse [28]. The resulting local optical absorption raises the temperature within the GST cell, whose magnitude depends on the presynaptic optical intensity, the postsynaptic feedback intensity, and the current PCM state. Consequently, synaptic weight updates occur only when sufficiently strong coincident pre- and post-synaptic optical activity is present, thereby implementing a local Hebbian learning rule directly through optical signal interactions.

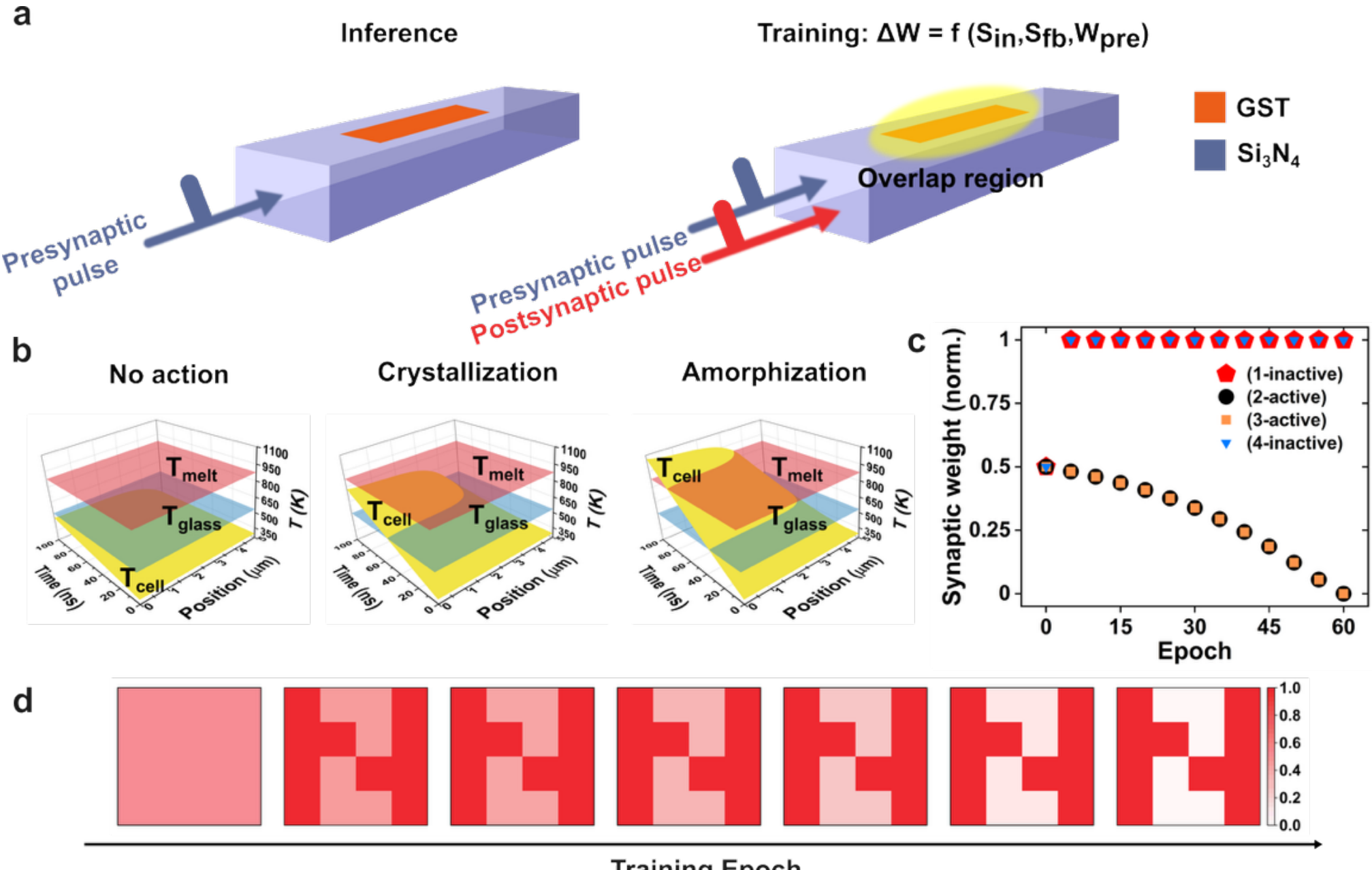

**Fig. 2** PCM synapse Hebbian learning dynamics. **(a)** Schematic illustration of the proposed PCM synapse learning mechanism. During learning, presynaptic input pulses (blue) and postsynaptic feedback pulses (red) overlap within the PCM-integrated waveguide crossing. The resulting local optical absorption induces temperature-dependent phase transitions within the GST cell, enabling synaptic weight adaptation according to a Hebbian learning rule. **(b)** Simulated spatiotemporal temperature evolution within the GST synapse during optical excitation. The blue and red planes indicate the glass-transition temperature $T_{glass}$ and melting temperature $T_{melt}$, respectively. When the local temperature remains below $T_{glass}$, no synaptic update occurs. Heating above $T_{glass}$ but below $T_{melt}$ induces crystallization and gradual synaptic-weight reduction, while heating above $T_{melt}$ produces amorphization and resets the synaptic weight to a high-transmission state. **(c)** Simulated evolution of representative synaptic weights corresponding to the first row of the input letter "N". The synapses are initialized in partially amorphous states and evolve through repeated presentations of the input pattern under local Hebbian feedback. **(d)** Simulated temporal evolution of the full synaptic-weight distribution during repeated presentation of the letter "N". Synapses associated with correlated pre- and post-synaptic activity progressively strengthen, while inactive synapses remain weak or gradually decay, enabling autonomous encoding of the input pattern through local Hebbian adaptation.

To model the synaptic weight evolution, the GST temperature evolution was modeled using a one-dimensional heat-diffusion equation commonly employed in PCM photonic thermal simulations: $\frac{\partial T(x,t)}{\partial t} = \alpha \frac{\partial^2 T(x,t)}{\partial x^2} + Q(x,t)$, where $T(x,t)$ denotes the local GST temperature distribution, $\alpha$ is the effective thermal diffusivity of the GST, and $Q(x,t)$ represents the optical absorption-induced heat source generated by the overlapping optical pulses [43]. The absorbed optical energy depends on the presynaptic input signal, the postsynaptic feedback signal, and the current PCM phase state, since the optical absorption coefficient varies between crystalline and amorphous GST. Consequently, the temperature evolution, and therefore the synaptic weight update, depends not only on the instantaneous optical excitation but also on the pre-existing synaptic state. Note that we employ a one-dimensional approximation here because heat transport is dominated along the waveguide propagation direction over the characteristic dimensions of the PCM interaction region.

Fig. 2b illustrates representative spatiotemporal temperature profiles within the GST synapse during optical excitation. Depending on the deposited optical energy and the initial phase distribution, different regions of the PCM cell experience distinct thermal histories, resulting in distinct synaptic-state transitions. The resulting weight evolution is therefore determined not only by the instantaneous optical excitation but also by the existing phase distribution within the GST. This history-dependent behavior supports gradual, state-dependent synaptic adaptation, enabling the PCM synapse to emulate key features of biological synaptic plasticity. The thermal thresholds shown in Fig. 2b are therefore used as criteria for determining synaptic-state evolution at each learning step.

To quantitatively model the evolution of the PCM synapse, we mapped normalized optical input signals to input pulse energies and normalized synaptic weights to the corresponding amorphous GST volume fraction. At each learning step, the thermal model evaluates whether the local optical excitation results in melting, crystallization, or no phase transition. If the local temperature exceeds $T_{melt}$, the melted GST region is converted into an amorphous state following rapid thermal quenching, thereby increasing the synaptic transmission weight. If the temperature remains between $T_{glass}$ and $T_{melt}$, crystal growth occurs at the crystalline-amorphous interface, reducing the amorphous volume fraction and decreasing the synaptic weight. A detailed flowchart of the synaptic update procedure and associated GST phase-transition dynamics is provided in Supplementary Information Section 1 and Fig. S1.

To illustrate the resulting Hebbian learning behavior, we simulated the unsupervised encoding of a pixelated letter "N" using the local optical feedback mechanism. During repeated presentations of the input pattern, each pixel's intensity was mapped to the initial transmission state of its corresponding PCM synapse. Synapses associated with correlated pre- and post-synaptic activity experienced repeated coincident optical excitation and progressively evolved toward stronger transmission states, while weakly correlated synapses underwent limited or no modification. Fig. 2c shows representative synaptic-weight trajectories for the four pixels in the first row of the letter "N," initialized from partially amorphous states. Fig. 2d presents the temporal evolution of the full synaptic weight distribution during repeated pattern presentation, demonstrating autonomous encoding of the input image through local Hebbian adaptation without externally computed gradients or labeled training data.

## 4. Experimental proof-of-concept platform

To experimentally validate the proposed PCM-based Hebbian learning framework, we implemented a PNN using commercially available fiber-optic components and real-time software control. The experimental platform incorporates the signal propagation, weighted optical summation, online synaptic adaptation, and local learning behavior of the proposed integrated photonic architecture while enabling flexible investigation of supervised and unsupervised learning dynamics under realistic hardware conditions.

Fig. 3 illustrates the experimental platform and its relationship to the proposed integrated deep PNN architecture. Optical weighted signal propagation and attenuation are physically implemented using optical fibers and variable optical attenuators (VOAs), while neuron activation functions and synaptic phase-transition dynamics are evaluated within a real-time software control loop. The experimentally implemented platform, therefore, serves as a proof-of-concept for the proposed PCM-based deep PNN, rather than a fabricated PIC.

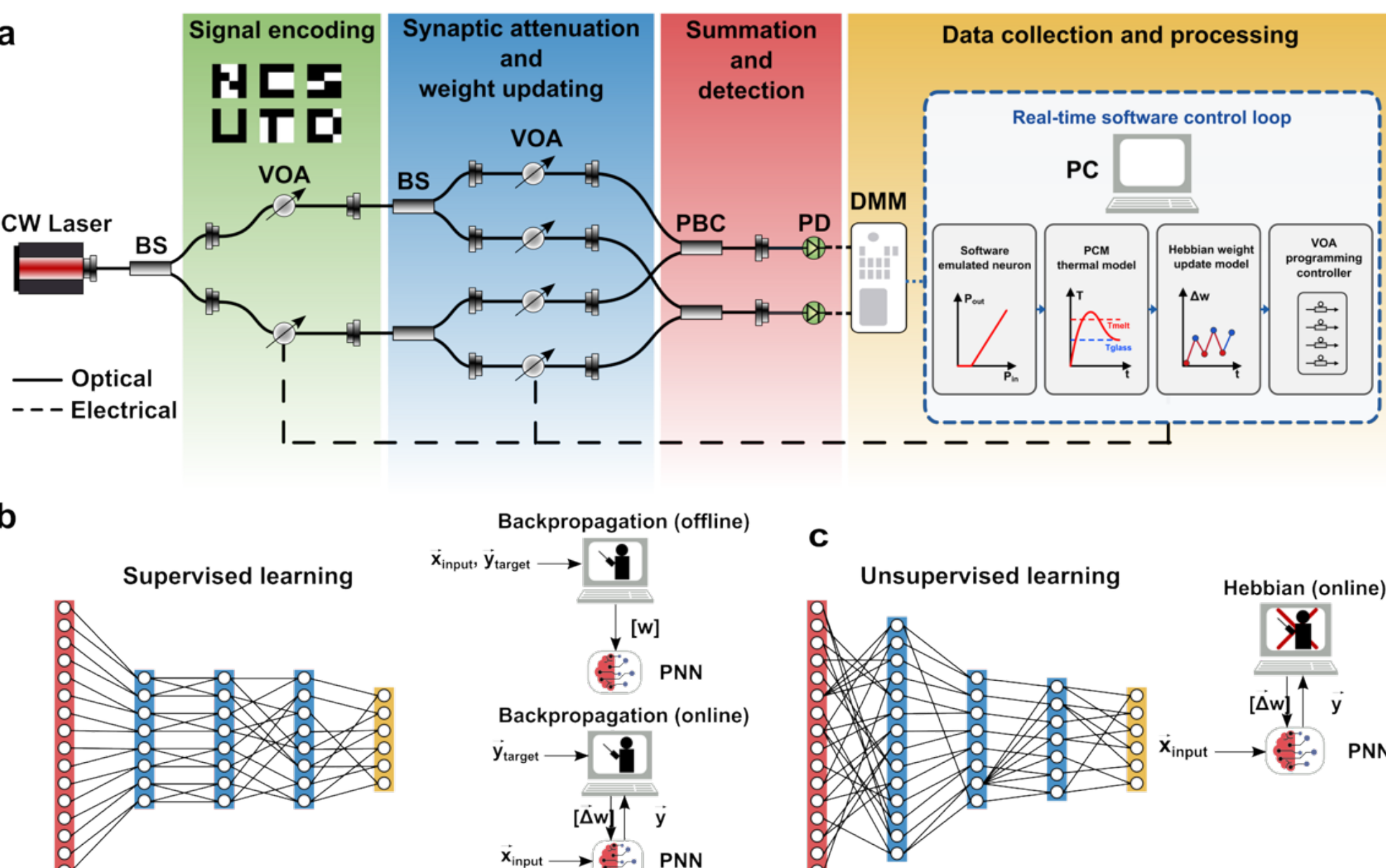

**Fig. 3** PNN experimental approach. **(a) An e**xperimental fiber-optic platform was used to validate the proposed PCM-based deep PNN learning framework. Optical-weighted signal propagation is implemented physically using fiber-optic components and programmable VOAs, while neuron activation functions and PCM synaptic evolution are evaluated within a real-time software control loop. The local optical feedback is functionally represented by photodetected postsynaptic signals that are routed through the real-time software control loop for Hebbian weight updating and VOA reprogramming. **(b)** PNN structure for supervised letter recognition. For offline backpropagation learning, the weights are pre-computed and programmed to the PNN (brain). For online backpropagation, the weights are continually updated using backpropagation on a PC. **(c)** PNN structure for unsupervised letter recognition. In the proposed fully integrated implementation, a local optical learning circuit could update the weights according to a Hebbian rule without PC-based backpropagation. In the present experimental setup, this process is implemented via the real-time software control loop shown in panel (a).

To clarify the relationship between the proposed integrated PCM architecture and the experimentally implemented proof-of-concept, Table 2 summarizes the correspondence between the major functional components of the two systems.

**Table 2:** Correspondence between the proposed PCM-based deep PNN architecture and the experimental proof-of-concept.

| Proposed integrated component | Experimental implementation |
|---|---|
| PCM synapse | VOA attenuation state |
| Optical weighted propagation | Fiber-optic links |
| PCM neuron | Software-emulated activation |
| Local optical feedback | Real-time feedback control loop |
| PCM phase transition dynamics | Thermal model in software |
| Synaptic memory state | VOA attenuation memory |
| Optical summation | Passive optical combining |

Although the VOAs do not inherently reproduce the physical phase-transition dynamics of PCM-based devices, the GST synaptic evolution model described in Section 3 was directly incorporated into the real-time experimental control loop. Specifically, the GST thermal phase-transition model, detailed in Supplementary Information Section 1, was used to determine the corresponding synaptic-state evolution during each learning step. The resulting synaptic transmission states were subsequently mapped to the VOA attenuation values, thereby enabling a preliminary experimental demonstration of the proposed PCM-based Hebbian learning process. Consequently, the experimental platform preserves the local learning dynamics, state-dependent synaptic evolution, and adaptive online learning behavior of the proposed architecture while utilizing experimentally accessible fiber-optic hardware.

In the proposed integrated architecture, the local optical feedback pathway physically routes a fraction of the postsynaptic optical signal back to the corresponding synapses within the same column. In this experiment, the local optical feedback process is represented by the measured post-synaptic optical output, acquired by the photodetectors and numerically routed within the real-time software control loop to the corresponding Hebbian synaptic update rule. Therefore, the platform preserves the functional locality of the Hebbian learning rule because each synaptic update depends only on the corresponding presynaptic input, the postsynaptic feedback signal, and the current synaptic state, although the feedback routing and PCM phase-transition dynamics are implemented in software rather than via an integrated optical feedback waveguide.

The experimental setup (Fig. 3a) consists of four primary functional blocks: optical signal encoding, programmable synaptic attenuation, optical summation and detection, and real-time software control. A continuous-wave laser operating at 1550 nm provides the optical carrier signal for the PNN. The optical input vector is encoded by a pair of VOAs that modulate the optical intensity based on the input pixel values. Beam splitters (BSs) distribute the optical signals into multiple weighted channels representing synaptic interconnections within the photonic crossbar of Fig. 1.

The programmable synaptic weights are implemented with a second set of VOAs that act as PCM synapses. Each VOA attenuation state corresponds to the transmission state of a PCM synapse determined by the GST Hebbian learning model described in Section 3. The weighted optical signals are subsequently combined using polarization beam combiners (PBCs) and detected using photodetectors (PDs), thereby implementing optical weighted summation analogous to the column-wise accumulation process in the proposed deep PNN architecture.

The detected optical signals are transferred to a real-time software control loop implemented using a data acquisition system and computer-based controller (PC). Within this software loop, the neuron activation function, PCM thermal evolution, Hebbian synaptic update, and VOA reprogramming operations are evaluated sequentially during each learning iteration. Specifically, the measured optical intensities are first processed using a software-emulated nonlinear neuron activation function. The resulting postsynaptic outputs are then used, together with the presynaptic optical intensities, to evaluate the PCM thermal model and determine the corresponding synaptic-state evolution according to the local Hebbian learning rule.

The updated synaptic transmission states are subsequently mapped onto the VOA attenuation values and reprogrammed into the optical network before the next learning iteration. Fig. 3a also summarizes the real-time learning workflow, including optical propagation, optical detection, evaluation of neuron activation, PCM thermal modeling, Hebbian weight updating, and VOA programming. This iterative feedback process enables real-time online adaptation of the experimentally implemented PNN while preserving the local learning behavior of the proposed PCM-based architecture. The modular architecture of the experimental setup enables flexible investigation of supervised learning (Fig. 3b), unsupervised Hebbian learning (Fig. 3c), and online adaptive learning dynamics in PNNs under realistic conditions of experimental noise, insertion loss, and device variability.

Using this photonic learning platform, we investigated both supervised and unsupervised learning behaviors under offline and online update conditions. Fig. 4 summarizes the resulting learning performance and adaptive synaptic evolution for the different learning configurations. The supervised learning experiments validate the trainability and inference capability of the experimental platform, while the unsupervised learning experiments demonstrate autonomous synaptic adaptation using the proposed local Hebbian learning framework.

### 4.1. Supervised learning and inference validation

We performed supervised learning experiments not as the main learning mechanism of the proposed architecture, but rather to validate the optical-weighted propagation, VOA-based weight programming, and the inference fidelity of the experimental platform. The supervised learning and inference capabilities of the experimental platform were evaluated using six pixelated alphabetic characters "NCSUTD" as representative input patterns. Both offline and online learning configurations were investigated to compare idealized weight programming and real-time adaptive learning behavior.

In the offline supervised learning configuration, synaptic weights were optimized digitally using conventional backpropagation prior to programming onto the VOA-based PCM synapses. During inference, optical input vectors were propagated through the weighted PNN, and optical weighted summation was performed through passive optical combining, followed by photodetection of the accumulated postsynaptic signals. Fig. 4a presents the experimentally measured output distributions for the six input characters under offline supervised inference conditions. Because the synaptic weights were preoptimized digitally prior to hardware implementation, the experimentally measured output states exhibit high classification fidelity.

To investigate adaptive learning behavior under realistic hardware conditions, we next implemented online supervised learning. In this configuration, experimentally measured optical signals were iteratively processed within the software control loop to evaluate neuronal activations, update synaptic weights, and reprogram the VOA-based synapses during each learning epoch. Unlike offline weight programming, the online learning process inherently incorporates experimental nonidealities, including laser power fluctuations, VOA repeatability limitations, insertion loss variations, detector noise, and polarization crosstalk.

Fig. 4b shows that despite accumulated experimental imperfections, the PNN successfully performed online supervised learning and converged toward the desired classification states through iterative adaptive weight updates. The close agreement between the offline and online supervised learning results demonstrates the robustness of the experimental learning framework and validates the PNN's ability to perform real-time adaptive photonic learning under experimentally realistic conditions.

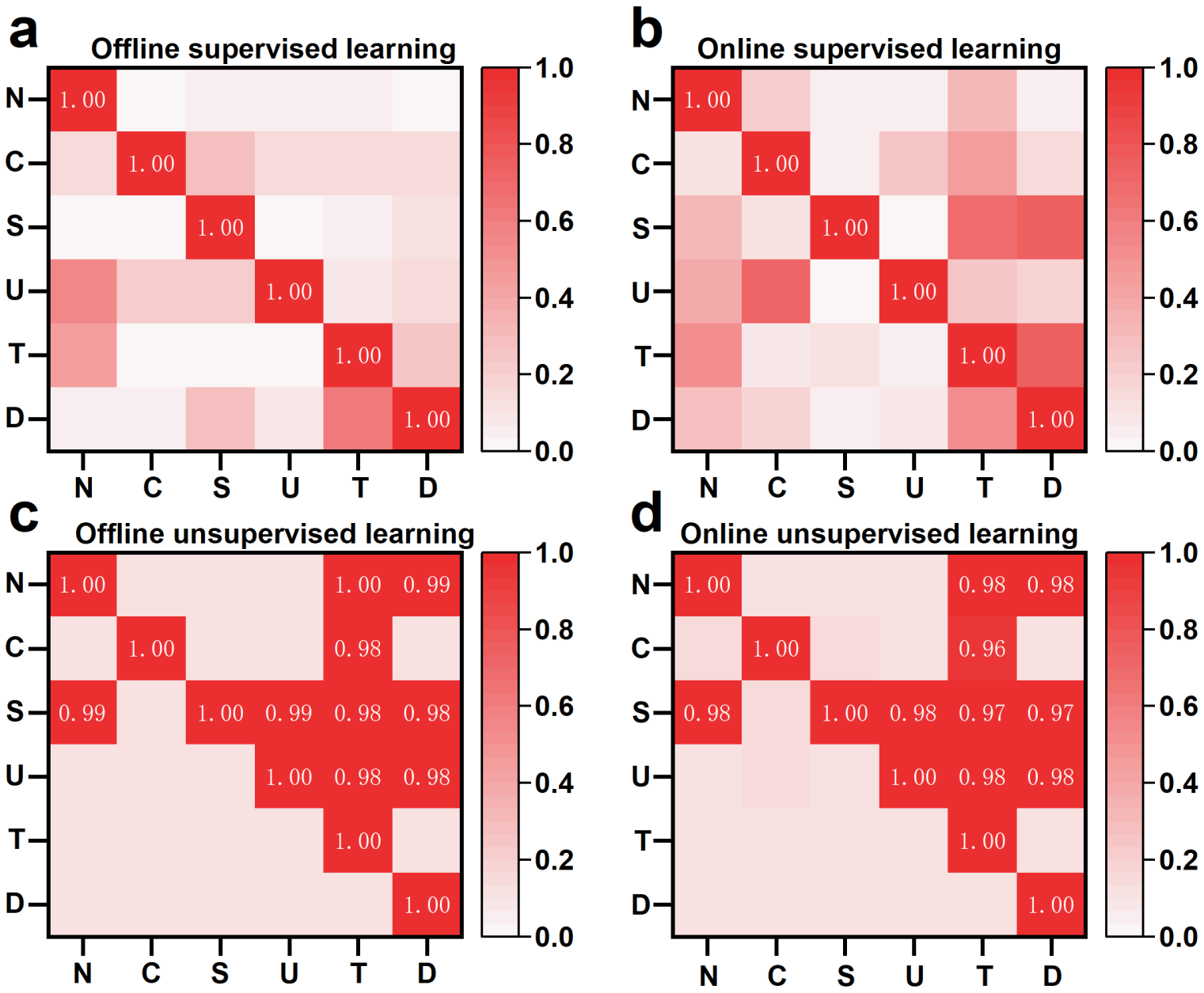

**Fig. 4** The experimental recognition results, shown via the normalized output power distribution after training. The input letters are labeled on the y-axis, and the output letters are labeled on the x-axis. In all four panels, the diagonal elements attain the highest values, indicating 100% recognition accuracy. Recognition was assigned according to the maximum output response for each input class. (a) Offline supervised learning; (b) Online supervised learning; (c) Offline unsupervised learning; (d) Online unsupervised learning.

### *4.2. Unsupervised Hebbian learning demonstration*

The proposed local Hebbian learning framework was also evaluated under both offline and online unsupervised learning conditions. Unlike supervised backpropagation-based learning approaches in Section 4.1, which require externally computed target gradients and labeled training data, the Hebbian learning mechanism relies exclusively on local correlations between pre- and post-synaptic optical activity to drive synaptic evolution.

For the offline unsupervised Hebbian-learning experiments, synaptic weight evolution was evaluated using the PCM Hebbian learning model described in Section 3, without real-time hardware feedback. Repeated presentations of the input patterns progressively modified the synaptic transmission states according to the local coincidence-driven update rule. Fig. 4c shows output values from the output layer after repeated unsupervised exposure to the input characters. The experimental synaptic states autonomously evolved to encode the dominant spatial features of the presented "NCSUTD" patterns without externally computed gradients or supervisory labels.

To further validate the proposed learning framework under realistic experimental conditions, we implemented online unsupervised Hebbian learning. In this configuration, experimentally measured optical signals were processed in real time within the software control loop to evaluate neuron activations, compute local Hebbian updates using the PCM thermal model, and iteratively reprogram the VOA-based synapses during each learning epoch. Fig. 4d shows that, compared with the offline Hebbian learning results of Fig. 4c, the online learning results exhibit increased distortions and non-uniformities, arising from accumulated hardware imperfections, including laser power fluctuations of ~0.02 dB, VOA repeatability limitations of ~0.1 dB, fiber insertion losses, detector noise, and polarization crosstalk. Nevertheless, the PNN successfully preserved the dominant spatial features of the input patterns through autonomous local synaptic adaptation. These results demonstrate the feasibility of implementing online, unsupervised Hebbian learning in PNN systems via local, optical feedback-driven synaptic evolution, without the need for globally computed gradients or externally supervised training signals.

## 5. Discussion

This work presents a PCM-based photonic neuromorphic learning framework that combines local Hebbian learning, non-volatile synaptic adaptation, and optical weighted signal propagation within a multilayer PNN architecture. By leveraging local coincidence-driven synaptic updates rather than globally computed gradients, the proposed approach provides a pathway toward scalable and energy-efficient photonic learning systems compatible with integrated photonic hardware.

Unlike many previously demonstrated PNN platforms that rely on offline digital training or externally computed backpropagation updates [14,19,21,25], the proposed framework employs local synaptic adaptation governed directly by the correlation between pre- and post-synaptic optical activity. This local learning mechanism is conceptually analogous to Hebbian plasticity in biological neural systems and eliminates the need for global error propagation during synaptic updating. In addition, the use of non-volatile PCM synapses enables persistent optical weight storage without static power consumption, offering advantages for future large-scale photonic neuromorphic hardware.

The experimental platform developed in this work enabled validation of both supervised and unsupervised learning behaviors under realistic experimental conditions. The supervised learning experiments demonstrated stable adaptive weight updates and successful optical inference, while the unsupervised learning experiments demonstrated autonomous synaptic evolution and pattern encoding via local Hebbian adaptation. Although the unsupervised learning results online exhibited distortions due to accumulated hardware imperfections, the dominant spatial features of the input patterns were successfully preserved through iterative adaptive learning.

Importantly, the present work provides experimental validation of the proposed PCM-based deep PNN learning framework rather than a fully integrated, all-optical chip-scale PNN implementation. In the current experimental platform, optical weighted signal propagation and summation were implemented physically using fiber-optic hardware, while neuron activation functions and PCM phase transition dynamics were evaluated within a real-time software control loop. Consequently, several important challenges remain before fully integrated photonic Hebbian learning systems can be realized experimentally.

First, large-scale integration of photonic crossbars will require low-loss photonic routing structures, compact waveguide crossings, including those with PCM waveguide synapses, and scalable optical fan-out architectures capable of supporting dense multilayer interconnectivity. Second, integrated optical PCM microring neuron implementations with low switching energy, high modulation contrast, and stable nonlinear activation behavior remain an active area of research. Third, thermal crosstalk, fabrication variability, accumulated optical loss, and device-to-device variability in PCM may significantly influence large-scale learning dynamics and network stability in future integrated implementations. In addition, practical PICs will require co-optimization of photonic device physics, optical signal distribution, thermal management, and learning algorithms to maintain stable online adaptive behavior under realistic operating conditions.

Despite these challenges, the proposed learning framework offers several attractive features for future photonic neuromorphic systems. Because synaptic updates depend only on locally available pre- and post-synaptic optical signals, the learning mechanism is inherently parallel and compatible with distributed photonic hardware implementations. Furthermore, the non-volatile nature of PCM synapses enables persistent analog optical weight storage without continuous power consumption, while the local optical-feedback architecture may reduce the need for repeated O-E-O conversions during adaptive learning. These characteristics may become increasingly important for future large-scale photonic AI hardware operating under strict energy-efficiency and latency constraints.

Future work will focus on the experimental realization of integrated PCM photonic synapses and neurons, the development of fully integrated optical-feedback pathways, the investigation of large-scale multilayer photonic learning dynamics, and the quantitative analysis of noise robustness and device variability in practical photonic neuromorphic systems. In addition, extending the proposed framework toward wavelength-division multiplexed architectures, spatiotemporal optical learning systems, and hybrid photonic-electronic neuromorphic platforms may further improve scalability and computational throughput.

Overall, the results presented here demonstrate the feasibility of local Hebbian learning in a PNN platform and lay the foundation for future integrated photonic neuromorphic systems capable of adaptive online learning via local optical feedback-driven synaptic evolution.

## 6. Methods

### *Measurement setup*

The experimental PNN proof-of-concept was implemented using commercially available fiber-optic components (Supplementary Fig. S3). Optical synapses were implemented with a 6-channel microelectromechanical systems (MEMS) VOA module from Dicon Fiberoptics. Two channels are dedicated to modulating the input signal amplitude, while the remaining four are configured in a 2×2 arrangement to adjust synaptic weights by controlling input signal attenuation. The MEMS VOAs, with their modest response time (< 2 ms) and high durability (> $1\times10^9$ cycles), enable real-time online learning for letter recognition tasks. Functioning as non-volatile, reconfigurable PCM optical synapses, these VOAs can be tuned via an RS-232 control interface, supporting both offline and online learning.

Optical neurons, which implement nonlinear activation functions, are simulated using software-based neuron models with inverse-sigmoid and ReLU-like activation functions. Due to phase differences between signals traveling through different optical paths, conventional beam combiners posed challenges in power combining. To mitigate such interference, polarized beam combiners are used to merge the signals before photodetection. A comprehensive analysis of output power stability and polarization crosstalk is provided in Section 4 of the Supplementary Information.

Given the constraints on available experimental resources, a 2×2 synapse unit was repeatedly reused to construct the PNN, with intermediate values from hidden layers temporarily stored in computer memory.

### *Supervised learning algorithm – the backpropagation method*

For supervised learning, we employ classical backpropagation based on gradient descent. The cost function, or error, measures the difference between the expected (Y) and actual ($O_a$) neuron output values:

$$E = Y - O_a$$

Whereas the gradient is calculated as a function of the errors:

$$dO = df(O) \times E$$

This gradient is then backpropagated through the network using the chain rule:

$$dC = df(C) \times (dO \times W4)$$

$$dB = df(B) \times (dC \times W3)$$

$$dA = df(A) \times (dB \times W2)$$

The weights for each layer of synapses are updated using the following rule:

$$W4 = \alpha \times W4 + \eta \times C_a \times dO$$

$$W3 = \alpha \times W3 + \eta \times B_a \times dC$$

$$W2 = \alpha \times W2 + \eta \times A_a \times dB$$

$$W1 = \alpha \times W1 + \eta \times X \times dA$$

where momentum $\alpha = 1$ and learning rate $\eta = 0.065$ are both constants. A, B, C, and O represent the output vectors for different hidden and output layers, while $A_a$, $B_a$, and $C_a$ represent the output vectors after the nonlinear activation has been applied.

### *Unsupervised learning algorithm – local Hebbian rule*

For unsupervised learning, we employ a  Hebbian rule. Synapse updates are based on the input signal, local feedback signal, and the current synaptic weight. Given the fixed timing relationship between input and output signals, synaptic weight updates depend solely on the combined amplitude of the input and feedback signals. Three distinct outcomes are possible: no action (weight maintenance), crystallization (weight decrease), and amorphization (weight increase). Unlike conventional backpropagation, which relies on global feedback, our Hebbian rule relies on local feedback. Additionally, it avoids differential computations, thereby eliminating the need for on-chip electronic components and making it ideally suited for all-optical neural network implementations. The weight update $\Delta w$ is a function of the i-th input presynaptic signal in the N-th layer $x_{i,N}$, the feedback signal in the N-th layer $y_N$, and the current synaptic weight $w$, $\Delta w = f(x_{i,N}, y_N, w)$.

**Acknowledgments**

This research was supported by two National Science Foundation (NSF) CAREER Awards (ECCS- 2209871 and CCF- 2146439), the Army Research Office (ARO) Young Investigator Program (W911NF-19-1-0303), and Defense Advanced Research Projects Agency (DARPA) MTO HR0011-25-3-0188.

**Author contributions**

X.L., J.S.F., and Q.G. conceived the PNN design. X.L. designed the experimental setup and performed all measurements with assistance from A.C. D.B. developed the training algorithm, with contributions from P.Z. and W.H.B. J.S.F. and Q.G. supervised the project. All authors participated in data analysis. X.L. wrote the manuscript with contributions from all authors.

**Competing financial interests**

The authors declare no competing interests.

**Additional information**

## References

1. S. R. Nandakumar, S. R. Kulkarni, A. V Babu, and B. Rajendran, "Building Brain-Inspired Computing Systems: Examining the Role of Nanoscale Devices," IEEE Nanotechnol. Mag. **12**, 19–35 (2018).
2. B. J. Shastri, A. N. Tait, T. de Lima, W. H. P. Pernice, H. Bhaskaran, C. D. Wright, and P. R. Prucnal, "Photonics for artificial intelligence and neuromorphic computing," Nat. Photonics **15**, 102–114 (2021).
3. C. D. Schuman, S. R. Kulkarni, M. Parsa, J. P. Mitchell, P. Date, and B. Kay, "Opportunities for neuromorphic computing algorithms and applications," Nat. Comput. Sci. **2**, 10–19 (2022).
4. D. Kudithipudi, C. Schuman, C. M. Vineyard, T. Pandit, C. Merkel, R. Kubendran, J. B. Aimone, G. Orchard, C. Mayr, R. Benosman, J. Hays, C. Young, C. Bartolozzi, A. Majumdar, S. G. Cardwell, M.

Payvand, S. Buckley, S. Kulkarni, H. A. Gonzalez, G. Cauwenberghs, C. S. Thakur, A. Subramoney, and S. Furber, "Neuromorphic computing at scale," Nature **637**, 801–812 (2025).

5. E. Painkras, L. A. Plana, J. Garside, S. Temple, F. Galluppi, C. Patterson, D. R. Lester, A. D. Brown, and S. B. Furber, "SpiNNaker: A 1-W 18-Core System-on-Chip for Massively-Parallel Neural Network Simulation," IEEE J. Solid-State Circuits **48**, 1943–1953 (2013).
6. F. Akopyan, J. Sawada, A. Cassidy, R. Alvarez-Icaza, J. Arthur, P. Merolla, N. Imam, Y. Nakamura, P. Datta, G.-J. Nam, B. Taba, M. Beakes, B. Brezzo, J. B. Kuang, R. Manohar, W. P. Risk, B. Jackson, and D. S. Modha, "TrueNorth: Design and Tool Flow of a 65 mW 1 Million Neuron Programmable Neurosynaptic Chip," IEEE Transactions on Computer-Aided Design of Integrated Circuits and Systems **34**, 1537–1557 (2015).
7. P. A. Merolla, J. V Arthur, R. Alvarez-Icaza, A. S. Cassidy, J. Sawada, F. Akopyan, B. L. Jackson, N. Imam, C. Guo, Y. Nakamura, B. Brezzo, I. Vo, S. K. Esser, R. Appuswamy, B. Taba, A. Amir, M. D. Flickner, W. P. Risk, R. Manohar, and D. S. Modha, "A million spiking-neuron integrated circuit with a scalable communication network and interface," Science (1979). **345**, 668–673 (2014).
8. M. Davies, N. Srinivasa, T.-H. Lin, G. Chinya, Y. Cao, S. H. Choday, G. Dimou, P. Joshi, N. Imam, S. Jain, Y. Liao, C.-K. Lin, A. Lines, R. Liu, D. Mathaikutty, S. McCoy, A. Paul, J. Tse, G. Venkataramanan, Y.-H. Weng, A. Wild, Y. Yang, and H. Wang, "Loihi: A Neuromorphic Manycore Processor with On-Chip Learning," IEEE Micro **38**, 82–99 (2018).
9. A. Neckar, S. Fok, B. V Benjamin, T. C. Stewart, N. N. Oza, A. R. Voelker, C. Eliasmith, R. Manohar, and K. Boahen, "Braindrop: A Mixed-Signal Neuromorphic Architecture With a Dynamical Systems-Based Programming Model," Proceedings of the IEEE **107**, 144–164 (2019).
10. J. Pei, L. Deng, S. Song, M. Zhao, Y. Zhang, S. Wu, G. Wang, Z. Zou, Z. Wu, W. He, F. Chen, N. Deng, S. Wu, Y. Wang, Y. Wu, Z. Yang, C. Ma, G. Li, W. Han, H. Li, H. Wu, R. Zhao, Y. Xie, and L. Shi, "Towards artificial general intelligence with hybrid Tianjic chip architecture," Nature **572**, 106–111 (2019).
11. W. Wan, R. Kubendran, C. Schaefer, S. B. Eryilmaz, W. Zhang, D. Wu, S. Deiss, P. Raina, H. Qian, B. Gao, S. Joshi, H. Wu, H.-S. P. Wong, and G. Cauwenberghs, "A compute-in-memory chip based on resistive random-access memory," Nature **608**, 504–512 (2022).
12. M. A. Nahmias, T. F. de Lima, A. N. Tait, H.-T. Peng, B. J. Shastri, and P. R. Prucnal, "Photonic Multiply-Accumulate Operations for Neural Networks," IEEE Journal of Selected Topics in Quantum Electronics **26**, 1–18 (2020).
13. J. Feldmann, N. Youngblood, M. Karpov, H. Gehring, X. Li, M. Stappers, M. Le Gallo, X. Fu, A. Lukashchuk, A. S. Raja, J. Liu, C. D. Wright, A. Sebastian, T. J. Kippenberg, W. H. P. Pernice, and H. Bhaskaran, "Parallel convolutional processing using an integrated photonic tensor core," Nature **589**, 52–58 (2021).
14. Y. Shen, N. C. Harris, S. Skirlo, M. Prabhu, T. Baehr-Jones, M. Hochberg, X. Sun, S. Zhao, H. Larochelle, D. Englund, and M. Soljačić, "Deep learning with coherent nanophotonic circuits," Nat. Photonics **11**, 441–446 (2017).
15. F. Ashtiani, A. J. Geers, and F. Aflatouni, "An on-chip photonic deep neural network for image classification," Nature **606**, 501–506 (2022).
16. X. Lin, Y. Rivenson, N. T. Yardimci, M. Veli, Y. Luo, M. Jarrahi, and A. Ozcan, "All-optical machine learning using diffractive deep neural networks," Science (1979). **361**, 1004–1008 (2018).
17. S. Behroozinia and Q. Gu, "Leveraging multiplexed metasurfaces for multi-task learning with all-optical diffractive processors," Nanophotonics **13**, 4505–4517 (2024).
18. W. Shi, Z. Huang, T. Fu, and H. Chen, "Review of nonlinear activation functions in optical neural networks," Advanced Photonics **7**, (2025).
19. J. Spall, X. Guo, and A. I. Lvovsky, "Training neural networks with end-to-end optical backpropagation," https://doi.org/10.1117/1.AP.7.1.016004 **7**, 016004 (2025).
20. L. Bernstein, A. Sludds, C. Panuski, S. Trajtenberg-Mills, R. Hamerly, and D. Englund, "Single-shot optical neural network," Sci. Adv. **9**, (2023).
21. L. G. Wright, T. Onodera, M. M. Stein, T. Wang, D. T. Schachter, Z. Hu, and P. L. McMahon, "Deep physical neural networks trained with backpropagation," Nature **601**, 549–555 (2022).
22. T. P. Lillicrap, D. Cownden, D. B. Tweed, and C. J. Akerman, "Random synaptic feedback weights support error backpropagation for deep learning," Nat. Commun. **7**, (2016).
23. M. J. Filipovich, Z. Guo, M. Al-Qadasi, B. A. Marquez, H. D. Morison, V. J. Sorger, P. R. Prucnal, S. Shekhar, and B. J. Shastri, "Silicon photonic architecture for training deep neural networks with direct feedback alignment," Optica **9**, 1323 (2022).
24. A. Momeni, H. Rajabalipanah, A. Abdolali, and K. Achouri, "Generalized Optical Signal Processing Based on Multioperator Metasurfaces Synthesized by Susceptibility Tensors," Phys. Rev. Appl. **11**, (2019).
25. A. Momeni, B. Rahmani, M. Malléjac, P. del Hougne, and R. Fleury, "Backpropagation-free training of deep physical neural networks," Science (1979). **382**, 1297–1304 (2023).
26. Y. Munakata and J. Pfaffly, "Hebbian learning and development," Dev. Sci. **7**, 141–148 (2004).
27. D. Querlioz, W. S. Zhao, P. Dollfus, J.-O. Klein, O. Bichler, and C. Gamrat, "Bioinspired networks with nanoscale memristive devices that combine the unsupervised and supervised learning approaches," in *Proceedings of the 2012 IEEE/ACM International Symposium on Nanoscale Architectures*, NANOARCH '12 (ACM, 2012), pp. 203–210.

28. P. Zhou, A. J. Edwards, F. B. Mancoff, S. Aggarwal, S. K. Heinrich-Barna, and J. S. Friedman, "Neuromorphic Hebbian learning with magnetic tunnel junction synapses," Communications Engineering **4**, (2025).
29. Z. Cheng, C. Ríos, W. H. P. Pernice, C. D. Wright, and H. Bhaskaran, "On-chip photonic synapse," Sci. Adv. **3**, (2017).
30. J. Feldmann, N. Youngblood, C. D. Wright, H. Bhaskaran, and W. H. P. Pernice, "All-optical spiking neurosynaptic networks with self-learning capabilities," Nature **569**, 208–214 (2019).
31. W. Zhou, B. Dong, N. Farmakidis, X. Li, N. Youngblood, K. Huang, Y. He, C. David Wright, W. H. P. Pernice, and H. Bhaskaran, "In-memory photonic dot-product engine with electrically programmable weight banks," Nat. Commun. **14**, (2023).
32. A. N. Tait, T. F. De Lima, E. Zhou, A. X. Wu, M. A. Nahmias, B. J. Shastri, and P. R. Prucnal, "Neuromorphic photonic networks using silicon photonic weight banks," Sci. Rep. (2017).
33. R. Hamerly, L. Bernstein, A. Sludds, M. Soljačić, and D. Englund, "Large-Scale Optical Neural Networks Based on Photoelectric Multiplication," Phys. Rev. X **9**, (2019).
34. M. Wuttig and N. Yamada, "Phase-change materials for rewriteable data storage," Nat. Mater. **6**, 824–832 (2007).
35. H.-S. P. Wong, S. Raoux, S. Kim, J. Liang, J. P. Reifenberg, B. Rajendran, M. Asheghi, and K. E. Goodson, "Phase Change Memory," Proceedings of the IEEE **98**, 2201–2227 (2010).
36. G. W. Burr, M. J. Breitwisch, M. Franceschini, D. Garetto, K. Gopalakrishnan, B. Jackson, B. Kurdi, C. Lam, L. A. Lastras, A. Padilla, B. Rajendran, S. Raoux, and R. S. Shenoy, "Phase change memory technology," Journal of Vacuum Science & Technology B, Nanotechnology and Microelectronics: Materials, Processing, Measurement, and Phenomena **28**, 223–262 (2010).
37. C. Rios, M. Stegmaier, P. Hosseini, D. Wang, T. Scherer, C. D. Wright, H. Bhaskaran, and W. H. P. Pernice, "Integrated all-photonic non-volatile multi-level memory," Nature Photonics 2015 9:11 **9**, 725–732 (2015).
38. M. Wuttig, H. Bhaskaran, and T. Taubner, "Phase-change materials for non-volatile photonic applications," Nat. Photonics **11**, 465–476 (2017).
39. C. Ríos, N. Youngblood, Z. Cheng, M. Le Gallo, W. H. P. Pernice, C. D. Wright, A. Sebastian, and H. Bhaskaran, "In-memory computing on a photonic platform," Sci. Adv. **5**, (2019).
40. J. Zheng, Z. Fang, C. Wu, S. Zhu, P. Xu, J. K. Doylend, S. Deshmukh, E. Pop, S. Dunham, M. Li, and A. Majumdar, "Nonvolatile Electrically Reconfigurable Integrated Photonic Switch Enabled by a Silicon PIN Diode Heater," Advanced Materials **32**, (2020).
41. J. S. Lee, N. Farmakidis, C. D. Wright, and H. Bhaskaran, "Polarization-selective reconfigurability in hybridized-active-dielectric nanowires," Sci. Adv. **8**, (2022).
42. C. Y. Lee, C. Lian, H. Sun, Y. S. Huang, N. Acharjee, I. Takeuchi, and C. A. Ríos Ocampo, "Spatial and temporal control of glassy-crystalline domains in optical phase change materials," Journal of the American Ceramic Society **107**, 1543–1556 (2024).
43. K. Aryana, H. J. Kim, C. C. Popescu, S. Vitale, H. Bin Bae, T. Lee, T. Gu, and J. Hu, "Toward Accurate Thermal Modeling of Phase Change Material-Based Photonic Devices," Small **19**, 2304145 (2023).